\documentclass[english]{bvm} 

\bvmdef\articlenumber{0000}

\bvmdef\type{V}

\date{}

\title{Quality monitoring of federated Covid-19 lesion segmentation}

\author{
	Camila Gonz\'{a}lez \inst{1}, 
	Christian~L. Harder \inst{1}, 
	Amin Ranem \inst{1},
	Ricarda Fischbach \inst{2}, 
	Isabel~J. Kaltenborn \inst{2},
	Armin Dadras \inst{2},
	Andreas~M. Bucher \inst{2}, 
	Anirban Mukhopadhyay \inst{1}
}

\authorrunning{Gonz\'{a}lez et al.}
\institute{
\textsuperscript{1} Medical and Environmental Computing, Technische Universität Darmstadt\\
\textsuperscript{2} Diagnostische und Interventionelle Radiologie, Universitätsklinikum Frankfurt
}

\email{camila.gonzalez@gris.tu-darmstadt.de}

\usepackage[colorlinks,allcolors=black,pdftex]{hyperref}
\usepackage{xcolor}
\usepackage{subfigure}
\usepackage{placeins}

\usepackage{array}
\usepackage{booktabs}

\begin{document}

%==============================================================================
% wählen Sie mit dem Befehl \selectlanguage die Sprache aus, in der Ihr 
% Proceeding verfasst ist
%
%\selectlanguage{ngerman}
\selectlanguage{english}

\maketitle

\begin{abstract}
Federated Learning is the most promising way to train robust Deep Learning models for the segmentation of Covid-19-related findings in chest CTs. By learning in a decentralized fashion, heterogeneous data can be leveraged from a variety of sources and acquisition protocols whilst ensuring patient privacy. It is, however, crucial to continuously monitor the performance of the model. Yet when it comes to the segmentation of diffuse lung lesions, a quick visual inspection is not enough to assess the quality, and thorough monitoring of all network outputs by expert radiologists is not feasible. In this work, we present an array of lightweight metrics that can be calculated locally in each hospital and then aggregated for central monitoring of a federated system. Our linear model detects over 70\% of low-quality segmentations on an out-of-distribution dataset and thus reliably signals a decline in model performance.
\end{abstract}

\section{Introduction}

The Covid-19 pandemic has strained medical resources across the world while demonstrating the value of time-saving workflow enhancements. Deep Learning solutions for the quantification of clinically relevant infection parameters, which segment Covid-19-characteristic lesions in CTs, have shown promising results.

Yet sufficient maturity for clinical use is frequently not reached by present approaches \cite{0000-01}. This is mainly due to neural networks failing silently coupled with a lack of appropriate quality controls. Scanner models and acquisition protocols vary between and within hospitals, changing image distribution. This causes deep learning models to produce low-quality outputs with high confidence \cite{0000-02}.

Covid-19-related ground glass opacities and consolidations can occur in various forms, from covering multiple small regions to diffuse affection of the entire lung \cite{0000-04}. Identifying low-quality segmentation masks is very time consuming and requires extensive experience, but thorough monitoring of all network outputs by expert readers is not logistically feasible.

Automated quality assurance for segmentation masks is not yet a developed field. Existing approaches include the training of a CNN on the logits of the segmentation prediction \cite{0000-10} or the concept of a Reverse Classification Algorithm \cite{0000-09} to predict segmentation quality. These are either computationally expensive or depend on rigid target shapes, which is not given in the case of Covid-19 lesions. Failed segmentations can however be identified by observing certain properties in the segmentation masks.

\begin{SCfigure}[5][t]
\label{0000-fig-01}
\setlength{\figbreite}{0.7\linewidth}
\caption{Quality features are extracted and an SVM model is used to perform inference locally at several hospitals. These quality scores are aggregated for each site and visualized at a central dashboard. In the entire process, only the privacy-preserving aggregated scores leave the institutions.} 
\includegraphics[width=\figbreite,height=0.7\figbreite]{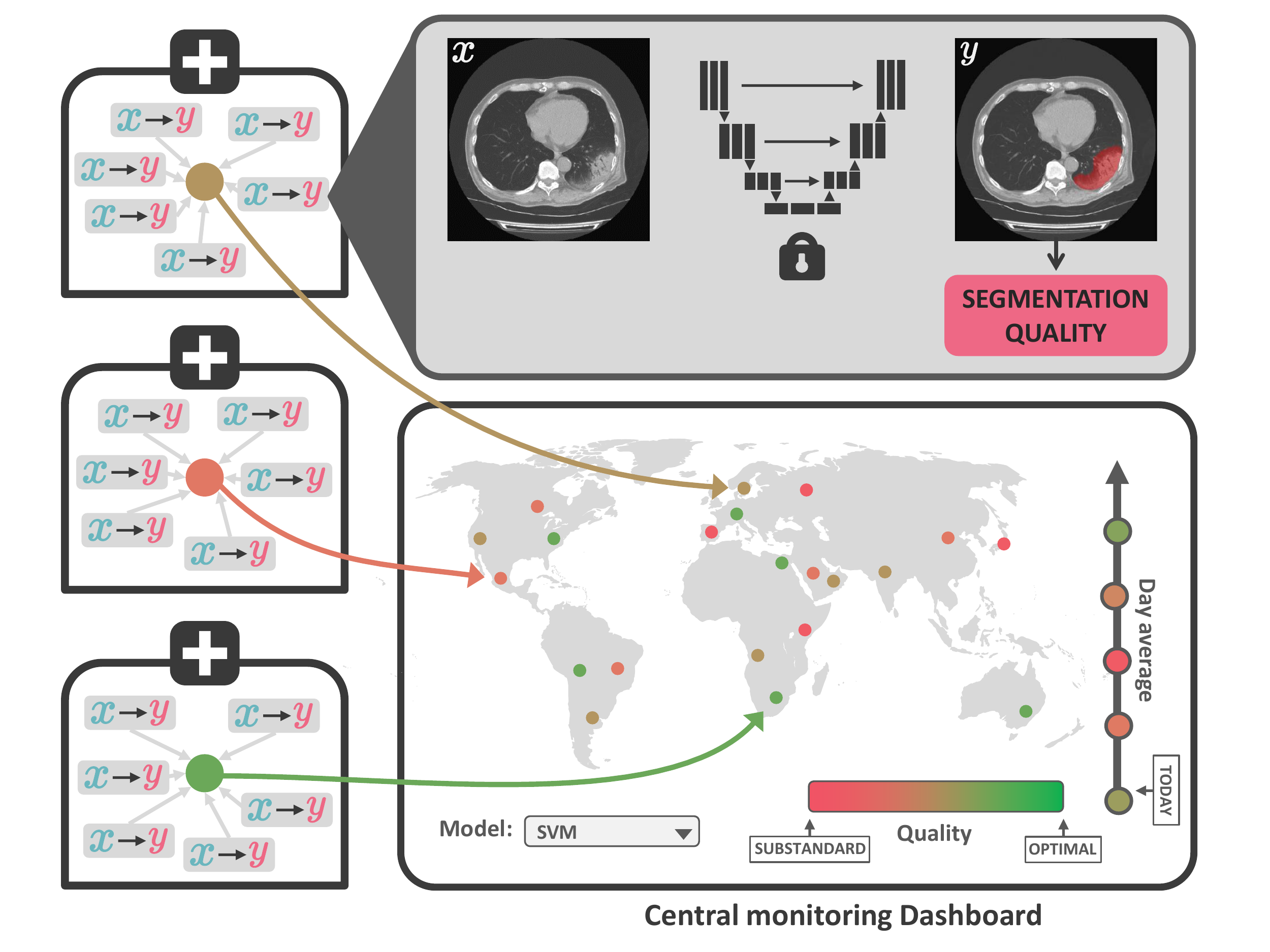}
\end{SCfigure}

We propose an array of \textbf{lightweight yet reliable quality metrics for segmentation masks that do not require ground truth annotations}. These can be \textbf{calculated locally without the need for expert reader review and then aggregated for each hospital for central monitoring of federated systems}, as illustrated in Fig. \ref{0000-fig-01}.

\section{Materials and Methods}

We implemented our code with Python 3.8 and PyTorch 1.6 and performed a retrospective study using several open-source datasets, as well as in-house data. The code can be found at github.com/MECLabTUDA/QA\_Seg.

\textbf{Data:} To obtain a dataset of predicted segmentations, we extracted predictions from an nnU-Net \cite{0000-03} trained on the COVID-19 Lung Lesion Segmentation Challenge (\textit{Challenge}) dataset \cite{0000-07}. We also predicted segmentations on \textit{MosMed} \cite{0000-08}, as well as in-house data with further 50 cases. Images were interpolated to dimension (50,512,512). Further details can be found in Table \ref{0000-tab-1}. We partitioned the predictions into in-distribution (ID) for the Challenge and in-house datasets (with which we trained our classifiers) and out-of-distribution (OOD) for MosMed. The ID datasets were randomly divided into \textit{ID train} and \textit{ID test}. We considered the Dice between ground truth and predicted masks as a measure of segmentation quality, as it is the most-used metric for segmentation overlap. As shown in Table \ref{0000-tab-1}, the ID data is heavily skewed towards good-quality segmentations. We define a \textit{failed} segmentation as having a Dice lower than 0.6 (following Valindria et al. \cite{0000-09}) and report their prevalence in Table \ref{0000-tab-1}.

\begin{table}[t]
\caption{Data distribution, including ratio of infection within the segmented lung volume \cite{0000-14}, nnU-Net performance and number of failed segmentation masks.}
\label{0000-tab-1}
\begin{tabular*}{\textwidth}{p{3cm}p{3cm}p{3cm}p{3cm}}
\hline
Property & Challenge  & In-house & MosMed \\ \hline
\textbf{Nr. cases (train, test)} & 199 (160, 39) & 50 (40, 10) & 50 (0, 50) \\
\textbf{Mean resolution} & (68.87,512.0,512.0) & (266.64,819.20,825.68) & (40.98,512.00,512.00) \\
\textbf{Infection ratio} & 0.061 $\pm$ 0.093 & 0.275 $\pm$ 0.274 & 0.016 $\pm$ 0.015 \\
\textbf{nnU-Net Dice (train) } & 0.75 $\pm$ 0.14 & 0.59 $\pm$ 0.2 & N.A. \\
\textbf{nnU-Net Dice (test) } & 0.71 $\pm$ 0.18 & 0.68 $\pm$ 0.1 & 0.47 $\pm$ 0.19 \\
\textbf{Failed masks (train)} & 24 & 12 & N.A. \\
\textbf{Failed masks (test)} & 8 & 1 & 37 \\
\end{tabular*}
\end{table}

\begin{figure}[ht]
    \centering
    \includegraphics[width=\textwidth]{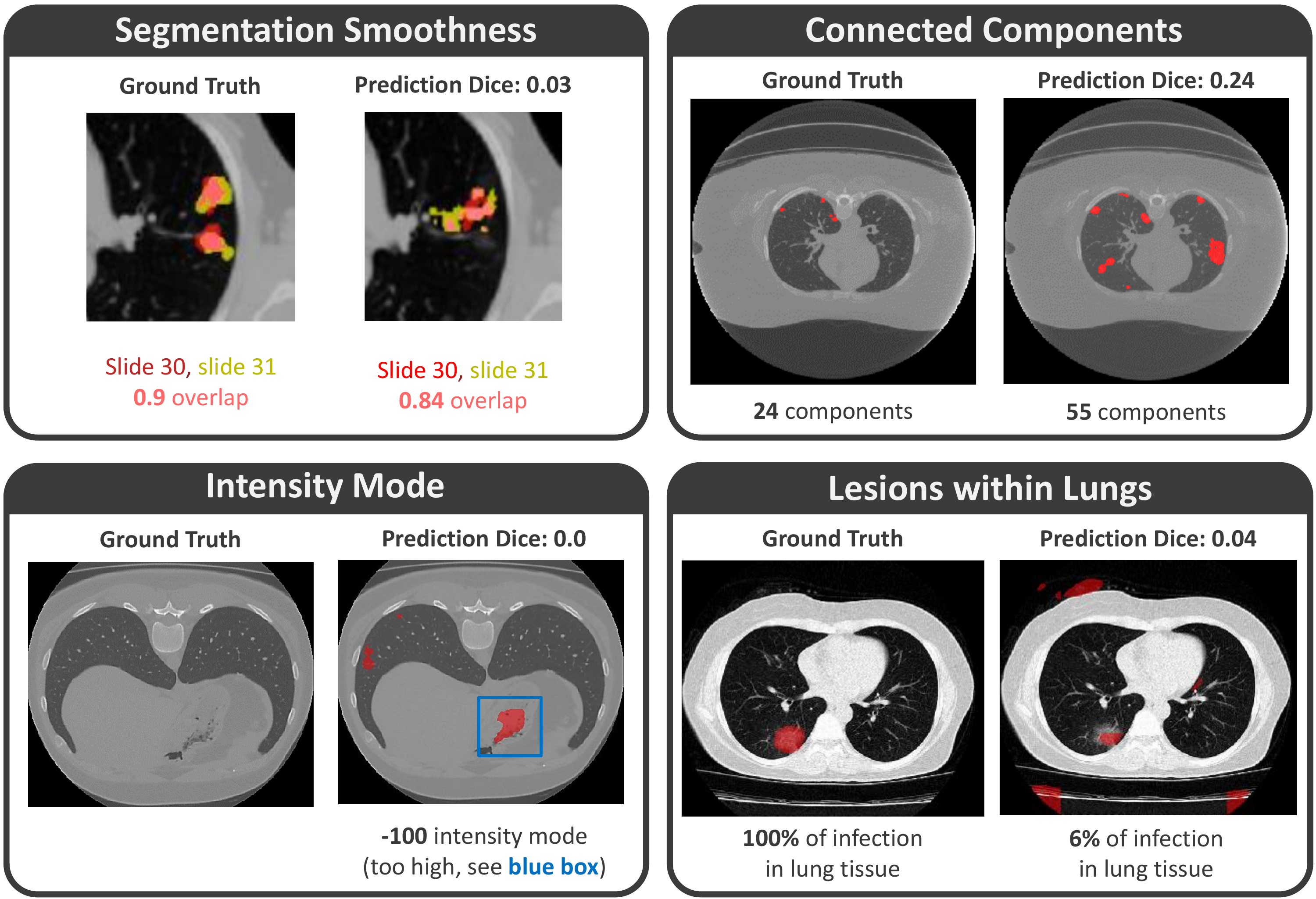}
    \caption{Exemplary subjects and slides for the four features used to assess segmentation quality.}
    \label{0000-fig-02}
\end{figure}

\textbf{Proposed features:} Inspired by van Rikxoort et al. \cite{0000-11}, we looked to predict the quality of segmentation masks - in the form of Dice coefficient - using only four features (see Fig. \ref{0000-fig-02}), defined as follows:

\begin{itemize}
\item \textit{Connected Components}: While lung lesions may occupy several components, failed segmentations are often more disconnected. We counted the number of connected components using Scikit-Image \cite{0000-12}, defining a component as one with a maximal distance of 3 by the City Block Metric to other voxels.
\item \textit{Intensity Mode}: Observing the intensity values in the CT, we can identify tissue that is very unlikely to be infected. Inspired by Kalka et al. \cite{0000-13}, we fitted a Gaussian distribution over the largest component and returned its mean.
\item \textit{Segmentation Smoothness}: In a correct segmentation mask, we expect two consecutive slices to have a high overlap and thus a high two-dimensional Dice. We computed the smoothness for every component by taking the average Dice scores for all consecutive slices that were not identical. We then averaged the smoothness over all components.
\item \textit{Lesions within Lungs}: A correct segmentation mask should be completely contained within the lung. To factor this in, we used a pre-trained lung segmentation model \cite{0000-14} and recorded the percentage of segmented tissue that is inside of the lung.  
\end{itemize}

\textbf{Models and training:} With these features, we trained and evaluated several models to predict the segmentation quality. We directly regressed the quality with a Ridge Regression (RR) and a Support Vector Regression (SVR) (trained until convergence) as well as a Multi-Layer-Perceptron (MLP) with (50,100,100,50) layers for 200 epochs minimizing the Mean Squared Error. We also discretized the quality values into five bins and performed classification with Support Vector Machine (SVM) and Logistic Regression (LR) models, using balanced class weights. Unless otherwise stated, we used the default Scikit-learn \cite{0000-15} library implementations.

\textbf{Evaluation:} As we were primarily interested in detecting failed segmentations, we report the sensitivity of all 5 models on this task. We also report the specificities for identifying the correct quality interval (averaged over 5 bins) on all ID and OOD datasets. In addition, we report the Mean Absolute Error as a metric than quantifies the ability of all models to directly predict the segmentation quality.

\section{Results}

In terms of sensitivity (detection of faulty segmentations) the classifiers (LR and SVM) outperformed the regression models by a large margin (see Table \ref{0000-tab-2}). This can be attributed to the class weights of the LR and SVM models balancing the disparately appearing classes in the training data, which improved their performance on differently distributed data. Though we were unable to detect the single failed segmentation out of 10 on the in-house dataset, we highlight the performance of the LR model, which detects over 60\% of failed segmentations on both of the bigger Challenge and MosMed datasets. All models showed a high specificity of over 0.8 on all datasets. The regression models achieved a lower mean absolute error but seemed to overfit the good-quality segmentations on the training dataset, which might explain their worse sensitivity.

We further evaluated the LR model using 10000 bootstrapping runs, sampling 192 data points from the training set and evaluating the model's sensitivity trained on these samples on the ID and OOD datasets for every run. We achieved 95\% confidence intervals for the sensitivity covering a range from 0.22 to 1.0. Furthermore, using a p-valued test with a significance level of 0.05, we can reject every null hypothesis stating that the sensitivity of the LR model is below 0.28.

In order to evaluate the individual contribution of each feature, we performed an ablation study where we left out each of the features for LR models. The "Intensity Mode" feature proved to be the least useful. Leaving it out allows us to correctly identify 5 more high-quality segmentations as such, though 9 faulty segmentations less are detected. All in all, using all four features achieves the best sensitivity-to-specificity trade-off.

We attribute most of the falsely classified segmentations to the low representation of bad segmentations in the training data and to these displaying plausible shapes. For example, segmentation masks covering only a few spots of healthy lung tissue, containing intensity values of possibly infected areas, while maintaining a smooth shape, were not detected.

\begin{table}[t]
\caption{Sensitivity of finding failed segmentations (Dice < 0.6), specificity of identifying the correct quality interval (avg. over 5 bins) and Mean Absolute Error  (mean+/- std) results for each model for ID and OOD datasets.}
\label{0000-tab-2}
\begin{tabular*}{\textwidth}{c@{\extracolsep\fill}cccccc}
\hline
\multicolumn{2}{c}{} & \multicolumn{2}{c}{Classifiers} & \multicolumn{3}{c}{Regressors} \\ \hline
\multicolumn{2}{c}{} & LR & SVM & RR & SVR & MLP \\ \hline
\multirow{3}{*}{Sensitivity} & Challenge & 0.63 (5/8) & 0.38 (3/8) & 0.38 (3/8) & 0.13 (1/8) & 0.25 (2/8) \\
 & In-house & 0.0 (0/1) & 0.0 (0/1) & 0.0 (0/1) & 0.0 (0/1) & 0.0 (0/1) \\
 & MosMed & 0.76 (28/37) & 0.68 (25/37) & 0.14 (5/37) & 0.35 (13/37) & 0.35 (13/37) \\
\hline
\multirow{3}{*}{Specificity} & Challenge & 0.84 & 0.85 & 0.88 & 0.87 & 0.87 \\
 & In-house & 0.8 & 0.83 & 0.95 & 0.9 & 0.9 \\
 & MosMed & 0.8 & 0.83 & 0.82 & 0.84 & 0.85 \\
\hline
\multirow{3}{*}{MAE
} & Challenge & 0.29 $\pm$ 0.22 & 0.26 $\pm$ 0.22 & 0.1 $\pm$ 0.1  & 0.11 $\pm$ 0.11  & 0.18 $\pm$ 0.13  \\
 & In-house & 0.24 $\pm$ 0.12 & 0.26 $\pm$ 0.14 & 0.08 $\pm$ 0.09  & 0.1 $\pm$ 0.07  & 0.09 $\pm$ 0.07  \\
 & MosMed & 0.33 $\pm$ 0.19 & 0.29 $\pm$ 0.23 & 0.22 $\pm$ 0.16  & 0.21 $\pm$ 0.18  & 0.23 $\pm$ 0.18  \\
\hline
\end{tabular*}
\end{table}

\section{Discussion}

We introduced a simple method to monitor performance of an nnU-Net trained to detect lung infections onset by Covid-19. We designed four features and found that a LR model using these reliably detects faulty segmentation masks. All the features are lightweight and do not require ground truth annotations, and so they can be used to monitor the deployment of a distributed, federated learning system.

Our findings have some limitations. First, we tested our methods retrospectively on a statically trained nnU-Net. This allowed us to accurately evaluate our methods, as we had access to ground truth test annotations, but a prospective study on a federated system with a few participating institutions would better emulate real deployment.

Secondly, the CT data was acquired on ICU patients, thus introducing considerable bias in patient demographics which are likely not representative of the general Covid-19 population. This also suggests that a measure other than Dice may be better suited for the general population, as the expressiveness of Dice is heavily dependent on lesion size.

Finally, each dataset was annotated by a different group of experts, so the definitions of the findings may vary across datasets. This is often the case when evaluating with OOD data but should be taken into account when considering the differences in performance.

In conclusion, training models in a federated fashion allows to leverage heterogeneous data sources without compromising patient privacy. However, it is necessary to constantly monitor the quality of the model outputs. In this work, we introduced an array of \textbf{lightweight quality metrics that can be calculated locally and aggregated for central monitoring}. These are particularly well-suited to the use case of lung lesion segmentation in chest CTs, as lesions vary greatly in terms of form and location and verifying their correctness is time-intensive even for trained radiologists. Future work should expand the metric catalogue and assess the effectiveness of the proposed methods in a model deployed across multiple hospitals. Our results present a first step towards an effective quality control of federated lung lesion segmentation.

\bibliographystyle{bvm}

\bibliography{3272}

%Kontrollzeile
\marginpar{\color{white}E\articlenumber}

\end{document}